\documentclass[dvips]{arxstspdf}
\usepackage{graphicx,flushend,stfloats}

\volume{25}
\issue{2}
\pubyear{2010}
\firstpage{245}
\lastpage{257}
\doi{10.1214/10-STS317}


\begin{document}
\begin{frontmatter}

\title{Approximate Dynamic Programming and Its Applications to the
Design of Phase I Cancer Trials}
\runtitle{Approximate Dynamic Programming and Phase I Cancer Trials}

\begin{aug}
\author[a]{\fnms{Jay} \snm{Bartroff}\corref{}\ead[label=e1]{bartroff@usc.edu}}
\and
\author[b]{\fnms{Tze Leung} \snm{Lai}\ead[label=e2]{lait@stanford.edu}}
\runauthor{J.~Bartroff and T. L.~Lai}

\address[a]{Jay Bartroff is Assistant Professor, Department of
Mathematics, University of Southern California, 3620 South Vermont
Ave, KAP 108, Los Angeles, CA 90089, USA \printead{e1}.}

\address[b]{Tze Leung Lai is Professor, Department of Statistics, and by
courtesy, Department of Health Research and Policy and Institute of
Computational and Mathematical Engineering, Stanford
University, Sequoia Hall, 390 Serra Mall, Stanford, CA 94305, USA
\printead{e2}.}

\end{aug}

%
\begin{abstract}
Optimal design of a Phase~I cancer trial can be formulated as a
stochastic optimization problem.
By making use of recent advances in approximate dynamic programming to
tackle the problem, we develop
an approximation of the Bayesian optimal design. The resulting design
is a convex combination of a
``treatment'' design, such as Babb et al.'s (\citeyear{Babb98}) escalation with
overdose control, and a
``learning'' design, such as Haines et al.'s (\citeyear{Haines03}) $c$-optimal design,
thus directly
addressing the treatment versus experimentation dilemma inherent in
Phase~I trials and providing a simple
and intuitive design for clinical use. Computational details are given
and the proposed design is compared
to existing designs in a simulation study. The design can also be
readily modified to include a first stage
that cautiously escalates doses similarly to traditional nonparametric
step-up/down schemes, while validating
the Bayesian parametric model for the efficient model-based design in
the second stage.
\end{abstract}

%

\end{frontmatter}

\section{Introduction}\label{sec1}

In typical Phase~I studies in the development of relatively benign
drugs, the drug is initiated at low doses and subsequently escalated to show
safety at a level where some positive response occurs, and healthy
volunteers are used as study subjects. This paradigm does not work for
diseases like cancer, for which a non-negligible probability of severe
toxic reaction has to be accepted to give the patient some chance of a
favorable response to the treatment. Moreover, in many such
situations, the benefits of a new therapy may not be known for a long
time after enrollment, but toxicities manifest
themselves in a relatively short time period. Therefore, patients
(rather than healthy volunteers) are used as study subjects, and given
the hoped-for
(rather than observed) benefit for them, one aims at an acceptable
level of toxic response in determining the dose. Current designs for
Phase~I cancer trials, which are sequential in nature, are an ad hoc
attempt to reconcile the objective of finding a \textit{maximum tolerated
dose} (MTD) with stringent ethical demands for protecting the study
subjects from toxicities in excess of what they can tolerate. It
treats groups of three patients sequentially, starting with the
smallest of an ordered set of doses. Escalation occurs if no toxicity
is observed in all three patients; otherwise an additional three
patients are treated at the same dose level. If only one of the six
patients has toxicity, escalation again continues; otherwise the trial
stops, with the lower dose declared as MTD. As pointed out by
Storer (\citeyear{Storer89}), these designs, commonly referred to as 3-plus-3
designs, are difficult to analyze, since even a
strict quantitative definition of MTD is lacking, ``although it should
be taken to mean some percentile of a tolerance distribution with
respect to some objective definition of clinical toxicity,'' and the
``implicitly intended'' percentile seems to be the 33rd percentile
(related to 2$/$6). Storer (\citeyear{Storer89}) also considered three other
``up-and-down'' sequential designs for quantile estimation in the
bioassay literature and performed simulation\break studies of their
performance in estimating the 33rd percentile. Subsequent simulation
studies by\break O'Quigley et~al. (\citeyear{OQuigley90}) showed the performance of these designs
to be
``dismal,'' for which they provided the following explanation: ``Not
only do (these designs) not make efficient use of accumulated data,
they make use of no such data at all, beyond say the previous three,
or sometimes six, responses.'' They proposed an alternative design,
called the \textit{continual reassessment method} (CRM), which uses
parametric modeling of the dose--response relationship and a Bayesian
approach to estimate the MTD or, more generally, the dose
level $x$ such that the probability $F(x)$ of a toxic event is $p$
($1/3$ in the case of MTD).

Letting $\theta=(\alpha,\beta)'$ and assuming the usual logistic model
%
\begin{equation}\label{1}
F_\theta(x)=1/\bigl\{1+e^{-(\alpha+\beta x)}\bigr\}
\end{equation}
for the probability of a toxic response at dose level~$x$, the problem
of optimal choice of $n$ dose levels to estimate the MTD seems to be
covered by the theory of nonlinear designs. A well-known difficulty in
nonlinear design theory is that the optimal design for parameter
estimation involves the unknown parameter vector. To circumvent the
difficulty, it has been proposed that the design be constructed
sequentially, using observations made to date to estimate $\theta$ by
maximum likelihood and choosing the next design point by using the MLE
to replace the unknown parameter value in the optimal design; see
Fedorov (\citeyear{Fedorov72}). If $\theta$ is known, then a target probability
$p$ of response is attained at the level $x_\theta$ that solves
$F_\theta(x_\theta) = p$, that is,
$x_\theta=[\log(p/(1-p))-\alpha]/\beta$. Wu (\citeyear{Wu85})
proposed to
use at stage $t + 1$ the \textit{certainty equivalence} (or plug-in)
level $x_{\hat\theta_t}$, where $\widehat\theta_t$ is the MLE of
$\theta$ based on $(x_i, y_i)$, $1 \leq i \leq t$, and $y_i$ is the
binary response at dose level $x_i$. Using some approximations, he
also derived a recursive representation of $x_{\hat\theta_t}$ and
showed that it is asymptotically equivalent (as $t \rightarrow
\infty$) to the adaptive stochastic approximation rule of Lai and Robbins (\citeyear{Lai79}).
The likelihood version of CRM proposed by O'Quigley and Shen (\citeyear{OQuigley96}) in
response to
the comments of Korn et~al. (\citeyear{Korn94}) on Bayesian designs is in fact a variant
of Wu's (\citeyear{Wu85}) design. Babb et~al. (\citeyear{Babb98}) pointed out
that the
symmetric nature of squared error loss used by CRM may not be
appropriate for modeling the toxic
response to a cancer treatment. They proposed the \textit{escalation
with overdose control} (EWOC) method, which uses an asymmetric linear
loss function that penalizes dose
level $x=\mathrm{MTD}+\delta$ for $\delta>0$, corresponding to an
overdose, more than an under-dose $x=\mathrm{MTD}-\delta$. Whereas CRM
is equivalent to estimating the MTD at each stage by the mean of the
posterior distribution of $x_\theta$, EWOC is
equivalent to estimating the MTD at each stage by the $\omega$th
quantile of the posterior distribution of $x_\theta$, where
$\omega\in(0,1/2)$ is the so-called \textit{feasibility bound},
usually chosen to be slightly less than $p$. There has also been much
work on designs
intended to give an accurate post-experiment estimate of the
$\mathrm{MTD}$ or other functions of the unknown parameter vector
$\theta$. For example, locally optimal designs such as $c$- and
$D$-optimal designs have been investigated extensively for binary
responses, and because a nonlinear model's
information matrix for binary data is a function of the unknown
parameters, locally optimal designs are usually applied using initial
estimates, multistage methods or Bayesian priors. Haines, Perevozskaya
and Rosenberger (\citeyear{Haines03})
proposed a two-stage design whose first stage is a locally optimal
design based on a chosen prior, which is then updated sequentially
during its second stage. A comparative study of these methods is given
in Section~\ref{sec:2stage}.

Since the parameter $\theta$ is unknown, \textit{active} statistical
learning involves setting the doses at levels that give maximum
information about the function of the unknown parameters of interest,
the MTD, and how to do this is a problem in nonlinear experimental
design theory (Abdelbasit and Plackett, \citeyear{Abdelbasit83}; Dette et~al., \citeyear{Dette04}). On the other
hand, there
is also an ethical issue of
treating patients in the Phase~I trial at dose levels below the
unknown MTD for safety, and hopefully close to the MTD for efficacy.
This dilemma between treatment of current patients and efficient
experimentation to gather information for future patients was
articulated by Lai and Robbins (\citeyear{Lai79}) in a simple linear regression model
$y_k=\alpha+\beta x_k+\varepsilon_k$, where, instead of the MTD, the
desired level is $(y^*-\alpha)/\beta$, for some given value $y^*$.
Whereas an asymptotic theory of how this dilemma can be resolved
optimally as $n\to\infty$ was developed by Lai and Robbins (\citeyear{Lai79}), it
was quite
recent that a tractable scheme was developed by Han, Lai and
Spivakovsky (\citeyear{Han06}) to
compute an approximately optimal solution for finite sample size $n$
(number of patients enrolled in the trial).

In Section~\ref{sec2} we introduce a basic stochastic optimization problem
that incorporates the treatment versus experimentation dilemma in the design
of Phase~I cancer trials. This problem adopts a Bayesian formulation
as in CRM and EWOC, for which the computation of
the posterior distributions of the parameters and of the MTD is
described in Section~\ref{sec2}. Because
the regression function $F_\theta(x)=E_\theta(y|x)$ for the binary
response $y$ given by (\ref{1}) is nonlinear in the parameters, the
stochastic optimization problem is considerably more difficult than
the linear regression model $E(y|x)=\alpha+\beta x$ considered by
Lai and Robbins (\citeyear{Lai79}). We review in Section~\ref{sec2} recent advances in the
field of
\textit{approximate dynamic programming}, which we use in Section~\ref{sec:hyb_des}
to develop a new tool for tackling the stochastic optimization problem.
Using this tool, we derive nearly optimal \textit{hybrid
designs} in Section~\ref{sec:hyb_des}. These hybrid designs are convex combinations
(and therefore hybrids) of designs that are
targeted toward treating the current patient at the best guess of the
MTD (e.g., EWOC and CRM) and the Haines--Perevozskaya--Rosenberger
designs that are $D$- or
$c$-optimal in estimating the model parameters for future
patients. The weights in these convex combinations are determined by
approximate dynamic programming and can be conveniently stored to
provide simple table look-up schemes for the
clinical user, as noted in Section~\ref{sec:conc} which gives some concluding
remarks. Section~\ref{sec:2stage} provides a comparative study of the hybrid
design and previous designs. It also introduces a modified hybrid
design that incorporates the traditional
nonparametric step-up/down scheme as a cautious first stage, followed
by the model-based design in the second stage of a Phase~I cancer trial.

\section{Stochastic Optimization Related to the Treatment
Versus Experimentation Dilemma}\label{sec2}

To begin with, we specify a prior distribution on $\theta$ by
following Babb et~al. (\citeyear{Babb98}) who first specify a range $[x_{\min},
x_{\max}]$ of possible dose values believed to contain the MTD, with
$x_{\min}$ believed to be a conservative starting value. Rather than
directly specifying the prior distribution $\pi$ for the unknown
parameter $\theta$ of the working model to be used in the second
stage, which may be hard for investigators to do in practice, an upper
bound~$q>0$ on the probability $\rho=F_\theta(x_{\min})$ of toxicity
at $x_{\min}$ can be elicited from investigators; uniform
distributions over $[x_{\min}, x_{\max}]$ and $[0,q]$ are then taken
as the prior distributions for the MTD and $F_\theta(x_{\min})$,
respectively. Let $\mathcal{F}_k$ denote the information set generated
by the first $k$ doses and responses, that is, by
$(x_1,y_1),\dots,(x_k,y_k)$. Letting $\eta$ denote the MTD, it is
convenient to transform from the
unknown parameters $(\alpha,\beta)$ in the two-parameter logistic
model (\ref{1}) to $(\rho,\eta)$ via the formulas\vspace*{-1pt}
%
\begin{eqnarray}
\label{2}\alpha&=&\frac{x_{\min}\log(1/p-1)-\eta\log(1/\rho
-1)}{\eta-x_{\min}},
\\[-2pt]
\label{3}\beta&=&\frac{\log(1/\rho-1)-\log(1/p-1)}{\eta-x_{\min}},\vspace*{-2pt}
\end{eqnarray}
giving\vspace*{-2pt}
\begin{eqnarray}\label{4}
\alpha+\beta x
 &=&\bigl((x-\eta)\log(1/\rho-1)\nonumber
 \\[-2pt]
 &&{}-(x-x_{\min
})\log(1/p-1)\bigr)\nonumber
\\[-8pt]\\[-8pt]
&&/(\eta-x_{\min})\nonumber
\\[-2pt]
&=&\psi(x,\rho,\eta).\nonumber\vspace*{-2pt}
\end{eqnarray}
Assuming that the joint prior distribution of $(\rho,\eta)$ has
density $\pi(\rho,\eta)$ with support on $[0,q]\times
[x_{\min},x_{\max}]$, the $\mathcal{F}_k$-posterior
distribution of $(\rho,\eta)$ has density\vspace*{-2pt}
\begin{eqnarray}\label{5}
\quad &&f(\rho,\eta|\mathcal{F}_k)\nonumber
\\[-2pt]
&&\quad =C^{-1}\prod_{i=1}^{k}\biggl[\frac
{1}{1+e^{-\psi
(x_i,\rho,\eta)}}\biggr]^{y_i}
\\[-2pt]
&&{}\qquad \hspace*{34pt}\cdot \biggl[\frac{1}{1+e^{\psi(x_i,\rho,\eta)}}\biggr]^{1-y_i}
\pi(\rho,\eta),\nonumber\vspace*{-2pt}
\end{eqnarray}
where\vspace*{-2pt}
\begin{eqnarray*}
C&=& \int_{x_{\min}}^{x_{\max}} \int_0^q
\prod_{i=1}^{k}\biggl[\frac{1}{1+e^{-\psi(x_i,\rho,\eta)}}\biggr]^{y_i}
\\[-2pt]
&&{}\hspace*{57pt}\cdot \biggl[\frac{1}{1+e^{\psi(x_i,\rho,\eta)}}\biggr]^{1-y_i}
\\[-2pt]
&&{}\hspace*{42pt}\times \pi
(\rho,\eta)
\,d\rho\, d\eta\vspace*{-2pt}
\end{eqnarray*}
is the normalizing constant. The marginal $\mathcal{F}_k$-posterior
distribution of $\eta$ is then\vspace*{-2pt}
%
\begin{equation}\label{6}
f(\eta|\mathcal{F}_k)=\int_0^q f(\rho,\eta|\mathcal{F}_k) \,d\rho.\vspace*{-1pt}
\end{equation}
The aforementioned CRM and EWOC doses based on $\mathcal{F}_k$ are the
mean and
the $\omega$-quantile of (\ref{6}).

\subsection{A Global Risk Function and Its Minimization}\label{sec:globrisk}

Note that using
EWOC or CRM amounts to the ``myopic'' policy of dosing the ($k+1$)th
patient at the dose $x_{k+1}=x$ that minimizes
$E[h(x,\eta)|\mathcal{F}_k]$, in which
%
\begin{equation}\label{7}
 h(x,\eta)=
\cases{
(x-\eta)^2 &for CRM,\cr
\omega(\eta-x)^+&\cr
\quad {}+(1-\omega)(x-\eta)^+&for EWOC,
}\hspace*{-23pt}
\end{equation}
where $x^+=\max(x,0)$ and
\[
E[h(x,\eta)|\mathcal{F}_k]=\int_{x_{\min}}^{x_{\max}} h(x,\eta)
f(\eta|\mathcal{F}_k)\,d\eta.
\]
Since the information about the
dose--toxicity relationship gained from $x_{k+1}$ and the response
$y_{k+1}$ affects the ability to safely and effectively dose the other
patients $k+2,k+3,\dots,n$, one potential weakness of these myopic
policies is that they may be inadequate in generating information on
$\theta$ for treating the rest
of the patients, as well as the post-experimental estimate of the MTD
for subsequent phases. To incorporate these considerations in a\break Phase~I
trial, $x_{1},x_{2},\dots,x_n$ should be chosen sequentially in such a
way as to minimize
the \textit{global risk}
%
\begin{equation}\label{8}
E\Biggl[\sum_{i=1}^n h(x_i,\eta)+g(\hat{\eta},\eta)\Biggr],
\end{equation}
in which the expectation is taken over the joint distribution of $(\rho
,\eta;x_1,y_1,\ldots,x_n,y_n)$.
Note that (\ref{8}) measures the effect of the dose $x_k$ on the
$k$th patient through $h(x_k,\eta)$, its effect on future patients in the
trial through~$\sum_{i=k+1}^n h(x_i,\eta)$, and its effect on the
post-trial estimate~$\hat{\eta}$ through $g(\hat{\eta},\eta)$. It
can therefore be used to address the dilemma between safe treatment of current
patients in the study and efficient experimentation to gather
information about $\eta$ for future patients. As noted in Section~\ref{sec1},
Lai and Robbins (\citeyear{Lai79}) have introduced a similar global risk function to address
the dilemma between information and control in the choice of $x_k$ in
the linear regression model $y_k=\alpha+\beta x_k+\varepsilon_k$ so
that the outputs~$y_k$, $1\leq k\leq n$, are as close as possible to
some target value $y^*$. Specifically, they consider (\ref{8}) with
$g=0$ and $h(x;\alpha,\beta)=(\alpha+\beta x -y^*)^2$.


Dynamic programming is a standard approach to a stochastic
optimization problem of the form~(\ref{8}). Define
%
\begin{equation}\label{9}
h_k(x)=
\cases{
E[h(x,\eta)|\mathcal{F}_k],
\qquad 0\leq k< n-1,
\cr
E[h(x,\eta)
\cr
{}\quad +g(\hat{\eta}(x_1,\ldots,x_{n-1},x),\eta
)|\mathcal{F}_{n-1}],
\cr
\qquad k=n-1.
}\hspace*{-20pt}
\end{equation}
To minimize (\ref{8}), dynamic programming solves for the optimal
design $x_1^*,\dots,x_n^*$ by backward induction that determines
$x_k^*$ by minimizing
%
\begin{equation}\label{10}
\quad\ h_{k-1}(x)+E\Biggl[\sum_{i=k+1}^n h_{i-1}(x_i^*)
\Big|\mathcal{F}
_{k-1},x_k=x\Biggr]
\end{equation}
after determining the future dose levels $x_{k+1}^*,\dots,x_n^*$. Note
that (\ref{10}) involves computing the conditional expectation of
$\sum_{i=k+1}^n h_{i-1}(x_i^*,\eta)$ given the dose $x$ at stage $k$ and
the information set $\mathcal{F}_{k-1}$, and that $x_k^*$ is
determined by
minimizing such conditional expectation over all $x$. For $i\geq k+1$,
since $x_i^*$ is a complicated nonlinear function of the past
observations and of $y_k,x_{k+1}^*,y_{k+1},\dots,x_{i-1}^*,y_{i-1}$
that are
not yet observed, evaluation of the aforementioned conditional
expectation is a formidable task. To overcome this difficulty, we use
recent advances in approximate dynamic programming, which we first
review and then extend and modify for the problem of minimizing the
global risk (\ref{8}).

\subsection{Rollout Algorithms}\label{sub31}

To begin with, consider the problem of minimizing (\ref{8}) with
$g=0$ and $h(x;\alpha,\beta)=(\alpha+\beta x -y^*)^2$ in the linear
regression model
$y_k=\alpha+\beta x_k+\varepsilon_k$ with i.i.d. normal errors
$\varepsilon_i$ having mean $0$. Assuming a normal prior distribution
of $(\alpha,\beta)$, the posterior distribution of $(\alpha,\beta)$ given
$\mathcal{F}_{i-1}$ is also bivariate normal with parameters
$E_{i-1}(\alpha), E_{i-1}(\beta),\break E_{i-1}(\alpha^2),
E_{i-1}(\beta^2), E_{i-1}(\alpha\cdot \beta)$, in which $E_{i-1}$ denotes
conditional expectation given $\mathcal{F}_{i-1}$. These conditional
moments have explicit recursive formulas; see Section 4 of
Han, Lai and Spivakovsky (\citeyear{Han06}). The myopic policy that
chooses $x$ at stage $i$ to
minimize $E[(\alpha+\beta x-y^*)^2|\mathcal{F}_{i-1}]$ is given
explicitly by
\begin{eqnarray}\label{11}
\quad \hat{x}_i&=&E_{i-1}\{(y^*-\alpha)\beta\}
/E_{i-1}(\beta^2)\nonumber
\\[-8pt]\\[-8pt]
&=&
\{y^*E_{i-1}(\beta)-E_{i-1}(\alpha\beta)\}
/E_{i-1}(\beta^2).\nonumber
\end{eqnarray}
Although the myopic policy is suboptimal for the global risk function
(\ref{8}), Han, Lai and Spivakovsky (\citeyear{Han06}) use it as a
substitute for the intractable $x_i^*$ for
$k+1\leq i\leq n$ in (\ref{10}), in which the conditional
expectation can then be evaluated by Monte Carlo simulation. This
method is called \textit{rollout} in approximate dynamic
programming. The idea is to approximate the
optimal policy $x_k^*$ by minimizing (\ref{10}) with
$x_{k+1}^*,\dots,x_n^*$ replaced by some known \textit{base policy}
$\hat{x}_{k+1},\dots,\hat{x}_n$, which ideally is some easily
computed policy
that is not far from the optimum. Specifically, given a base policy
$\hat{\mathbf{x}}=(\hat{x}_{1},\dots,\hat{x}_n)$, let $\hat
{x}_k^{(1)}$ be
the $x$ that minimizes
%
\begin{equation}\label{12}
\quad\ h_{k-1}(x)+E\Biggl[\sum_{i=k+1}^n
h_{i-1}(\hat{x}_i) \Big| \mathcal{F}_{k-1}, \hat{x}_k=x\Biggr],
\end{equation}
and the expectation in the second term in (\ref{12}) is typically
evaluated by Monte Carlo simulation. The policy $\hat{\mathbf{x}
}^{(1)}=(\hat{x}_1^{(1)},\dots,\hat{x}_n^{(1)})$ is
called the \textit{rollout} of $\hat{\mathbf{x}}$ and has been used for
stochastic control problems arising in a variety of applications; see
Section 2.1 of Han, Lai and Spivakovsky (\citeyear{Han06}). The
rollout $\hat{\mathbf{x}}^{(1)}$ may itself be used
as a base policy, yielding $\hat{\mathbf{x}}^{(2)}$, and, in theory, this
process may be repeated an arbitrary number of times, yielding
$\hat{\mathbf{x}}^{(1)}, \hat{\mathbf{x}}^{(2)}, \hat{\mathbf
{x}}^{(3)}, \ldots.$ Letting
$R(\mathbf{x})=E[\sum_{i=1}^n h_{i-1}(x_i)]$,
Bayard (\citeyear{Bayard91}) showed that, regardless of the base policy, rolling
out $n$ times yields the optimal design and that rolling out always
improves the base design, that is, that
\begin{eqnarray}\label{13}
R(\hat{\mathbf{x}})&\ge& R\bigl(\hat{\mathbf{x}}^{(1)}\bigr)\ge R\bigl(\hat{\mathbf
{x}}^{(2)}\bigr)\ge
\cdots\nonumber
\\[-8pt]\\[-8pt]
&\ge& R\bigl(\hat{\mathbf{x}}^{(n)}\bigr)=R(\mathbf{x}^*)\nonumber
\end{eqnarray}
for any policy $\hat{\mathbf{x}}$, where $\mathbf{x}^*$ denotes the
optimal policy.

For the global risk function (\ref{8}) associated with Phase~I
designs, with $h$ given by (\ref{7}), one can use the myopic design
EWOC or CRM as the base design in the rollout procedure. In contrast
with the explicit formula (\ref{11}) for the case of a linear
regression model with normal errors $\varepsilon_t$, the posterior
distribution with density function (\ref{5}) does not have
finite-dimensional sufficient statistics and the myopic design involves
(a) bivariate numerical integration to evaluate
\[
E[h_i(x_{i+1})|\mathcal{F}_{k-1}, x_k=x]
\]
for $i\ge k$, and (b) minimization of the conditional expectation over
$x$. The simulation studies in \ref{sec:2stage}, in which the rollout
is implemented with EWOC as the base design, show substantial
improvements of the rollout over EWOC and CRM. Although (\ref{13})
says that rolling out a base design can
improve it and rolling out $n$ times yields the dynamic programming
solution, in practice, it is difficult to use a rollout (which is
defined by a backward induction algorithm that involves Monte Carlo
simulations followed by numerical optimization at every stage) as the
base policy for another rollout. To overcome this
difficulty, we need a tractable representation of successive rollouts,
which we develop by using other ideas from approximate dynamic
programming (ADP).

\subsection{Combining Least Squares with Monte Carlo in~ADP}\label{sec:LSMC}

The conditional expectation in (\ref{10}), as a function of $x$, is
called the \textit{cost-to-go function} in dynamic programming. An ADP
method, which grew out of the machine learning (or, more specifically,
reinforcement learning) literature, is based on two statistical
concepts concerning the conditional expectation. First, for given $x$
and the past information $\mathcal{F}_{k-1}$, the conditional
expectation is an expectation and therefore can be evaluated by Monte
Carlo simulations, if one knows how $h_k(x_{k+1}^*),\ldots
,h_{n-1}(x_{n}^*)$ are generated. The second concept is that, by (\ref
{9}), $h_i(x_{i+1})$ is a conditional expectation given $\mathcal
{F}_{i}$, which is a regression function (or minimum-variance
prediction) of $h_i(x_{i+1})$, with regressors (or predictors)
generated from $\mathcal{F}_{i}$. Based on a large sample (generated
by Monte Carlo), the regression function can be estimated by least
squares using basis function approximations, as is typically done in
nonparametric regression. Combining least squares~(LS) regression with
Monte Carlo~(MC) simulations yields the following LS-MC method for
Markov decision problems in reinforcement learning. Let $\{s_t, t\ge0\}
$ be a Markov chain whose transition probabilities from state~$s_t$ to
$s_{t+1}$ depend on the action~$x_t$ at time~$t$, and let $f_t(s,x)$
denote the cost function at time~$t$, incurred when the state is $s$
and the \mbox{action~$x$} is taken. Consider the statistical decision problem
of choosing $x$ at each stage~$k$ to minimize the cost-to-go function
\begin{eqnarray}
\quad&& Q_k(s,x)\nonumber
\\
&&\quad=E\Biggl\{f_k(s,x)
\\
&&{}\qquad\quad\ +\sum_{t=k+1}^n f_t(s_t,x_t)\Big|
s_k=s, x_k=x\Biggr\},\nonumber
\end{eqnarray}
assuming that $x_{k+1},\ldots,x_n$ have been determined. Let
%
\begin{equation}\label{eq:argmin}
 V_k(s)=\min_x Q_k(s,x),\quad x_k^*=\arg\min_x Q_k(s,x).\hspace*{-25pt}
\end{equation}
These functions can be evaluated by the backward induction algorithm of
dynamic programming:\break $V_n(s)=\min_x f_n(s,x)$, and for $n>k\ge1$,
\begin{eqnarray}\label{eq:Bellmaneq}
\qquad V_k(s)&=&\min_x\{f_k(s,x)\nonumber
\\[-8pt]\\[-8pt]
&&{}\hspace*{19pt}+E[V_{k+1}(s_{k+1})|s_k=s, x_k=x]\},\nonumber
\end{eqnarray}
in which the minimizer yields $x_k^*$. The LS-MC method uses basis
functions~$\phi_j$, $1\le j\le J$, to approximate $V_{k+1}$ by
$\widehat{V}_{k+1}=\sum_{j=1}^J a_{k+1,j}\phi_j$, and uses this approximation
together with $B$ Monte Carlo simulations to approximate
\[
E[V_{k+1}(s_{k+1})|s_k=s, x_k=x]
\]
for every $x$ in a grid of representative values. This yields an
approximation~$\widetilde{V}_k$ to $V_k$ and also $\widehat{x}_k$ to
$x_k^*$. Moreover, using the sample
%
\begin{equation}\label{eq:MCdata}
\{(s_{k,b},\widetilde{V}_k(s_{k,b})), 1\le b\le B\}
\end{equation}
generated by the control action $\widehat{x}_k$, we can perform least
squares regression of $\widetilde{V}_k(s_{k,b})$ on $(\phi
_1(s_{k,b}),\ldots,\break \phi_J(s_{k,b}))$ to approximate $\widetilde{V}_k$
by $\widehat{V}_k=\sum_{j=1}^J a_{k,j}\phi_j$. Further details of this
approach can be found in Chapter~6 of Bertsekas (\citeyear{Bertsekas07}).

Although the problem (\ref{10}) can be viewed as a Markov decision
problem with the $\mathcal{F}_{t+1}$-posterior distribution being the
state $s_t$, the state space of the Markov chain at hand is
infinite-dimensional, consisting of all bivariate posterior distributions of the
unknown parameter vector $(\alpha,\beta)$. If the state space were
finite-dimensional, for example, $\mathbb{R}^m$, then one could
approximate the value functions (\ref{eq:argmin}) by commonly used
basis functions in nonparametric regression, such as regression splines
and their tensor products; see Hastie, Tibshirani and Friedman
(\citeyear{Hastie01}). However, in the infinite-dimensional case,
there is no such simple choice of basis functions of posterior
distributions, which are the states. As pointed out in Section~6.7 of
Bertsekas (\citeyear{Bertsekas07}), an alternative to approximating the value
functions $V_k$, called \textit{approximation in value space}, is to
approximate the optimal policy by a parametric family of policies so
that the total cost can be optimized over the parameter vector. This
approach is called \textit{approximation in policy space} and most of
its literature has focused on finite-state Markov decision problems and
gradient-type optimization methods that approximate the derivatives of
the costs, as functions of the parameter vector, by simulation. We now
describe a new method for approximation in policy space, which uses
iterated rollouts to optimize the parameters in a suitably chosen
parametric family of policies.

The choice of the family of policies should involve domain knowledge
and reflect the kind of policies that one would like to use for the
actual application. One would therefore start with a set of real-valued
basis functions of the state~$s_t$ of the Markov chain with general,
possibly infinitely-dimensional, state space, on which the family of
chosen policies will be based. The control policies in this family can
be represented by $\pi_t(\phi_1(s_t),\ldots, \phi_m(s_t);\bolds{\beta
})$, which is the action taken at time $t$ [after $s_t$ has been
observed and the basis functions $\phi_1(s_t),\ldots,\phi_m(s_t)$
have been evaluated] and in which $\bolds{\beta}$ is a parameter to be
chosen iteratively by using successive rollouts, with
\[
\bigl\{\pi_t\bigl(\phi_1(s_t),\ldots, \phi_m(s_t);\bolds{\beta}^{(j)}\bigr), 1\le
t\le n\bigr\}
\]
being the base policy for the rollout $\mathbf{x}^{(j+1)}$. Using the
simulated sample
\[
\bigl\{\bigl(s_{k,b},x_{k,b}^{(j+1)}\bigr), 1\le b\le B\bigr\},
\]
in which $s_{k,b}$ denotes the $b$th simulated replicate of $s_k$,
least squares regression of $x_{k,b}^{(j+1)}$ on $\pi_k(\phi
_1(s_{k,b}),\break \ldots,\phi_m(s_{k,b});\bolds{\beta})$ is performed to
estimate $\bolds{\beta}$ by $\bolds{\beta}^{(j+1)}$; nonlinear least
squares is used if $\pi_k$ is nonlinear in~$\bolds{\beta}$. In view of
(\ref{13}), each iteration is expected to provide improvements over
the preceding one. A concrete example of this method in a prototypical
Phase~I setting is given in the next section, where linear regression
splines are used in iterated rollouts.
In this setting the state variable~$s_t$ represents the complete
treatment history up to time~$t$ in the trial---all prior
distributions, doses and responses up to that time---and the cost
function $f_t(s_t,x)$ will be replaced by $h_t(x)$ given by (\ref{9}).

\section{Hybrid Designs as Base Policies for Iterated Rollouts}\label
{sec:hyb_des}

In their use of rollouts to approximate the optimum for (\ref{8}) for
the normal model, Han, Lai and Spivakovsky (\citeyear{Han06}), Section~3, used the structure of their problem to come up with an
ingenious ``perturbation of the myopic rule'' as a base policy to
improve the performance of the rollout, without performing second- or
higher-order rollouts. In this section we explore this technique in the
context of Phase~I designs, using such perturbations---called here
\textit{hybrid designs}---both as base policies and as a way to
represent highly complicated but efficient policies in a simple,
clinically useful way. As pointed out in Section~\ref{sec:globrisk},
the objective function of
the dynamic programming problem (\ref{8}) involves both
experimentation (for
estimating the MTD) and treatment (for the patients in the
study). Consider the $k$th patient in a trial of length $n$ ($\ge k$).
If the $k$th patient were the last patient to be treated in the trial
($n=k$), the best dose to give him/her would be the myopic dose $m_k$
that minimizes $h_{k-1}(x_k)$, given by (\ref{9}). On the other hand,
early on in the trial, especially if $n-k$ is relatively large, one
expects the optimal dose to be perturbed from $m_k$ in the direction of
a dose that provides more information about the dose--response model,
for the relatively large number of doses that will have to be set for
the future patients. Since the optimal design theory for learning the
MTD under overdose
constraints, developed by Haines, Perevozskaya and Rosenberger
(\citeyear{Haines03}), yields a $c$- or $D$-optimal
design $\ell_k$, we propose to use the following
\textit{hybrid design} representation of the optimal dose sequence:
%
\begin{equation}\label{14}
x_k^*=(1-\varepsilon_k)m_k+\varepsilon_k \ell_k,
\end{equation}
where $\ell_k$ is the chosen ``learning design.'' Of course, any
dosing policy admits the representation (\ref{14}) with
\[
\varepsilon_k=\frac{x_k^*-m_k}{\ell_k-m_k}\cdot\mathbf{1}_{\{\ell
_k\ne m_k\}}.
\]
However, we will show that it is possible to use rollouts to choose
$\varepsilon_k$ of a simple form, not depending on~$x_k^*$, such that the
resulting hybrid design given by the right-hand side of (\ref{14}) is
highly efficient. Similar ideas have been used in ``$\varepsilon$-greedy
policies'' in reinforcement learning (Sutton and Barto, \citeyear{Sutton98}, page~122).

From our simulation studies that include the example in Section~\ref
{sec:sim}, we have found that the sequential $c$-optimal design
(Haines et~al., \citeyear{Haines03}, Section 5) with~$c$ being the vector~$(0,1)'$
works well for
learning design $\ell_k$ in (\ref{14}), which we now briefly explain.
In general, optimal designs such as $c$- and $D$-optimal can be
characterized as optimizing some convex loss function $\Psi$ of the
information matrix~$I(\theta,\xi)$ associated with the parameter
value $\theta$ and a measure $\xi$ on the space of design points
(see Fedorov, \citeyear{Fedorov72}). Here $(nI(\theta,\xi))^{-1}$ is
interpreted as the asymptotic variance of the MLE $\widehat{\theta}_n$
of $\theta$. The optimization problem can be generalized to the
sequential Bayes setting, with prior distribution $\pi$ on $\theta$,
by finding the $\xi$ that minimizes
%
\begin{equation}\label{eq:infint}
\int\Psi[I(\theta,\xi_{k-1})+I(\theta,\xi)]\pi(\theta|\mathcal
{F}_{k-1})\,d\theta
\end{equation}
at the $k$th stage, where $\xi_{k-1}$ is the empirical measure of the
previous design points. In the case $k=1$, (\ref{eq:infint}) is
replaced by $\int\Psi[I(\theta,\xi)]\pi(\theta)\,d\theta$. For a
given vector~$c$, the $c$-optimal design measure $\xi$ minimizes the
asymptotic variance of the linear estimator $c'\widehat{\theta}_n$ of
$c' \theta$ or, equivalently, $\Psi[I(\theta,\xi)]=c'(I(\theta,\xi
))^{-1}c$. Taking the Bayesian $c$-optimal design with $c=(0,1)'$ as
the learning design $\ell_k$ in (\ref{14}) gives $c'\theta=c'(\alpha
,\beta)'=\beta$, hence, this design is optimal, in some sense, for
learning about $\beta$ or, equivalently, about the slope
\begin{eqnarray*}
\frac{\partial}{\partial x}E(y|x)\bigg|_{x=\eta}&=&
\frac{\partial}{\partial x}\biggl(\frac{1}{1+e^{-(\alpha+\beta
x)}}\biggr)\bigg|_{x=\eta}
\\
&=&\beta p(1-p)
\end{eqnarray*}
of the dose response curve (\ref{1}) at the MTD, for which $p$ is
$1/3$ or some other prespecified value. This has the following
connections to the stochastic optimization problem of Lai and Robbins
(\citeyear{Lai79})
discussed in Section~\ref{sec1} and to the rollout procedure of Han, Lai and
Spivakovsky (\citeyear{Han06}). For the normal model discussed in
Section~\ref{sec1} and as an asymptotic limiting case of other models, Sacks
(\citeyear{Sacks58}) showed that the optimal value of the step size (a
user-supplied parameter in the Lai--Robbins procedure affecting its
convergence rate) is proportional to $(\partial/\partial x)E(y|x)$.
Moreover, Han, Lai and Spivakovsky (\citeyear{Han06}), Section~3,
found that in the normal model, perturbations of the myopic policy in
the direction of this $c$-optimal design provide a base design for a
rollout that has comparable performance to that of an ``oracle policy.''

\subsection{Relating $\varepsilon_k$ to the Uncertainty in the Bayes
Estimate $E(\eta|\mathcal{F}_{k-1})$}\label{sec:4.1}

Since the treatment versus experimentation\break dilemma discussed in Section~\ref
{sec2} stems from the uncertainty in the current estimate of the MTD
$\eta$, it is natural to expect that the amount of perturbation from
the myopic dose $m_k$ depends on the degree of such uncertainty, using
little perturbation when the posterior distribution of $\eta$ is
peaked, and much more perturbation when it is spread out. This suggests
choosing $\varepsilon_k$ as a function of the posterior variance $\nu
_{k-1}^2=\operatorname{Var}(\eta|\mathcal{F}_{k-1})$, whose
reciprocal is called
the ``precision'' of $E(\eta|\mathcal{F}_{k-1})$ in Bayesian
parlance. Following the approach described in Section~\ref{sec:LSMC},
we use functions of $s_k=\nu_{k-1}/\nu_0$ as basic features of the
posterior distribution of $\eta$ to approximate the $\varepsilon_k$
in~(\ref{14}).

To begin, Monte Carlo simulations are performed to obtain the rollout
$\mathbf{x}^{(1)}$ of EWOC, yielding a simulated sample $\{
(e_{k,b},s_{k,b})$, $1\le b\le B$\}, where $e_{k,b}$ is the $b$th
simulated replicate of
%
\begin{equation}\label{eq:e_k}
e_k=\frac{x_k^{(1)}-m_k}{\ell_k-m_k} \cdot\mathbf{1}_{\{\ell_k\ne m_k\}},
\end{equation}
which is essentially the same as (\ref{14}) with $(x_k^*,\varepsilon_k)$
replaced by $(x_k^{(1)},e_k)$. The basic idea in Section \ref
{sec:LSMC} can be implemented via nonparametric regression of $e_{k,b}$
on $s_{k,b}$, yielding the estimated regression function $g_k$. Letting
$\widehat{e}_k=g_k(s_k)$, the hybrid design $x_k=(1-\widehat
{e}_k)m_k+\widehat{e}_k\ell_k$ can then be used as the base policy to
form the rollout
$\mathbf{x}^{(2)}$, and this procedure can be repeated to obtain the
iterated rollouts $\mathbf{x}^{(3)}, \mathbf{x}^{(4)},\ldots.$

Linear regression splines, and their tensor products for multivariate
regressors, provide a convenient choice of basis functions; see
Section~9.4 of Hastie, Tibshirani and Friedman (\citeyear{Hastie01}).
For the present problem, it suffices to use a truncated linear function
\begin{eqnarray}\label{eq:f1}
g_k(s)=\min\bigl\{1,\bigl(\beta_k^{(0)}+\beta_k^{(1)} s\bigr)^+\bigr\}\nonumber
\\[-8pt]\\[-8pt]
\eqntext{\mbox{for }
s_*\le s\le s^*,}
\end{eqnarray}
where $s_*$ and $s^*$ are the minimum and maximum of the sample
values~$s_{k,b}$, $1\le b\le B$, and to extend beyond the
range~$[s_*,s^*]$ by
%
\begin{equation}\label{eq:f12}
g_k(s)=\cases{
s g_k(s_*)/s_*, &$0\le s\le s_*$, \cr
g_k(s^*), &$s\ge s^*$,
}
\end{equation}
which agrees with the constraint $g_k(0)=0$ and ensures that the weight
assigned to experimentation does not exceed $g_k(s^*)$. A further
simplification is to group the data into $K$ blocks so that
$\varepsilon
_k=\varepsilon_k(s)$ does not vary with $k$ within each block, since
it is
expected that the amount of experimentation for the initial stages
depends mostly on the uncertainty about $\eta$, while for the final
stages experimentation would only benefit the post-trial estimate of
$\eta$.

\begin{table*}
\tablewidth=348pt
\caption{Risk, bias and RMSE of the final MTD estimate,
DLT rate and overdose rate (OD) of EWOC, rollout (ROLL) of EWOC, and
1st and 2nd hybrid approximations}\label{table:op-char}
\begin{tabular*}{348pt}{@{\extracolsep{4in minus 4in}}lccccc@{}}
\hline
\textbf{Design}&\textbf{Risk}&\textbf{Bias}&\textbf{RMSE}&\textbf{DLT}&\textbf{OD}\\
\hline
EWOC& 0.84 (0.01)& $-$0.20 (0.010) & 0.31 (0.04)&29.8\% (0.7\%) & 21.9\%
(0.6\%) \\
\\
ROLL&
0.75 (0.01) &$-$0.04 (0.009) & 0.22 (0.03) & 33.0\% (0.7\%) & 31.2\%
(0.7\%) \\
\\
Hybrid 1& 0.75 (0.02) &$-$0.14 (0.012) & 0.29 (0.06) & 33.5\% (1.5\%) &
37.5\% (1.5\%) \\
\\
Hybrid 2&0.71 (0.01) &$-$0.04 (0.005) & 0.22 (0.04) & 31.24\% (0.9\%) &
27.8\% (0.9\%)\\
\hline
\end{tabular*}
\end{table*}

\subsection{Example and Simulation Study}\label{sec:sim}

\begin{figure*} 

\includegraphics{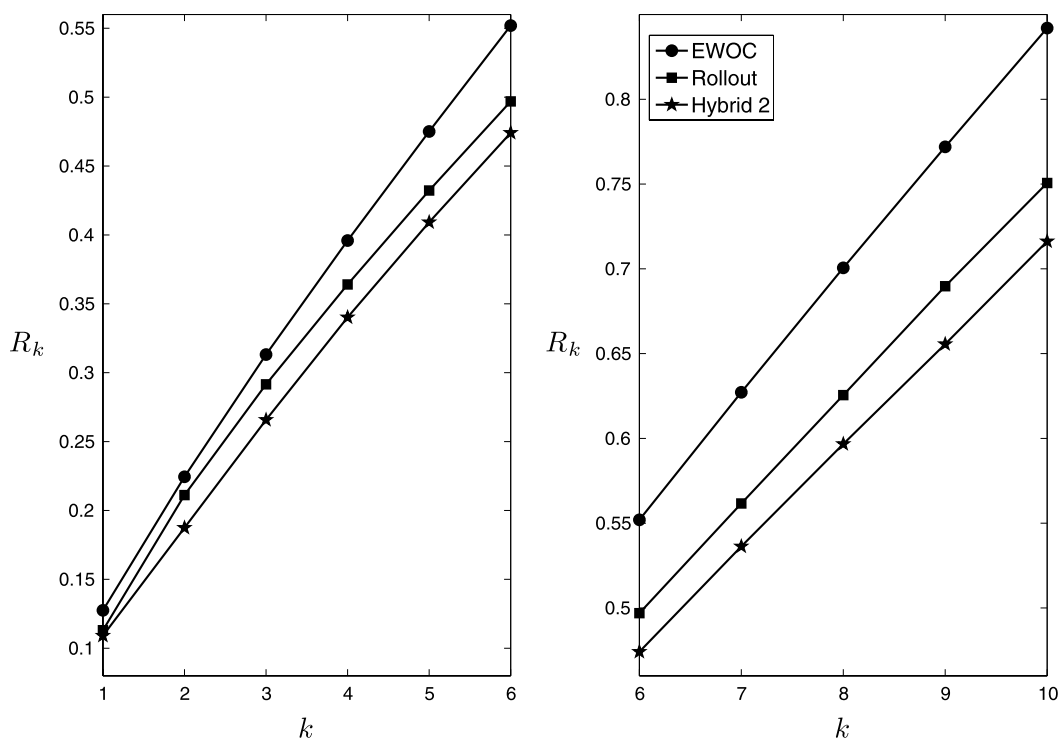}

\caption{Risk for EWOC, rollout of EWOC and hybrid designs.}
\label{fig:Rk}
\end{figure*}

We illustrate the method in Section \ref{sec:4.1} by applying it to
the following example, in which $n=10$ and $[x_{\min},x_{\max}]$ is
transformed to $[0,1]$ by location and scale changes. Independent
uniform priors on $[0,q]$ and $[0,1]$ are used for $\rho=F_\theta
(x_{\min})$ and the MTD $\eta$, respectively; see (\ref{2}) and
(\ref{3}) and the sentence following it. We use $q=1/3$ and the EWOC
loss with $\omega=1/4$ in (\ref{7}), and the squared error loss
$g(\widehat{\eta},\eta)=(\widehat{\eta}-\eta)^2$ in (\ref{8}). Since
$n$ is relatively small, we can assume for simplicity that $(\beta
_k^{(0)},\beta_k^{(1)})$ in (\ref{eq:f1}) does not vary with $k$ and
estimate the common $(\beta^{(0)},\beta^{(1)})$ by applying least
squares regression to the sample
\[
\{ (e_{k,b},s_{k,b})\dvtx 1\le k\le n, 1\le b\le B\}.
\]
We also simply use (\ref{eq:f1}) for all $s$ without performing the
extrapolation beyond $[s_*,s^*]$. Rolling out EWOC as the base design
and using $B=2000$ simulations, the preceding procedure gave $(\beta
^{(0)},\beta^{(1)})=(0.096,0.02)$. Putting
%
\begin{equation}\label{eq:epsk}
\varepsilon_k=\min\{1,(0.096+0.02\nu_{k-1}/\nu_0)^+\}
\end{equation}
in the hybrid design
%
\begin{equation}\label{eq:hyb_des}
x_k^{(1)}=(1-\varepsilon_k)m_k+\varepsilon_k \ell_k,
\end{equation}
we used $\mathbf{x}^{(1)}$ as the base policy of a second rollout, for
which the preceding procedure yielded $(\beta^{(0)},\break \beta
^{(1)})=(-0.72,0.94)$. Here we used the sequential $c$-optimal design
with $c=[0,1]'$ as the learning design $\ell_k$ (see Section~\ref
{sec:hyb_des}). Table \ref{table:op-char} contains the operating
characteristics, explained below, of EWOC and its rollout, the first
hybrid design $\mathbf{x}^{(1)}$ with $\varepsilon_k$ given by (\ref{eq:epsk})
and the second hybrid design $\mathbf{x}^{(2)}$ in which $(0.096,0.02)$ in
(\ref{eq:epsk}) is replaced by $(-0.72,0.94)$. Each result is based on
2000 simulation runs. The values of $(\rho,\eta)$ were generated
from the prior distribution given by the joint uniform distribution on
$[0,q]\times[x_{\min},x_{\max}]$. Figure \ref{fig:Rk} plots the cumulative
risk $R_k=\sum_{i=1}^k E[h_{i-1}(x_i)]$ of the EWOC, rollout and
hybrid designs for $k=1,\ldots, n(=10)$. The operating characteristics
in Table \ref{table:op-char} are the Monte Carlo estimates of overall risk
$R_{10}$, the bias and root mean squared error (RMSE) of the terminal
MTD estimate $\widehat{\eta}_{10}$, the DLT rate $P(y=1)$ and the
overdose rate OD, which is the expected proportion of patients treated
at doses higher than $\eta$. Standard errors are given in parentheses.

The first hybrid design, which is an approximation to the rollout
design, provides more than 10\% improvement in terminal risk $R_{10}$
over the myopic policy. The second hybrid design provides an additional
5\% improvement in the terminal risk $R_{10}$, and also smaller values
of the DLT and OD rates than the rollout design. The Monte Carlo
simulations used to evaluate the operating characteristics and to fit
the hybrid designs were performed by using rejection sampling to
simulate from the posterior distribution. At each stage, the posterior
distribution of $(\rho,\eta)$ is continuous and supported on the
compact set $[0,q]\times[x_{\min},x_{\max}]$, hence, the joint
uniform distribution on $[0,q]\times[x_{\min},x_{\max}]$ is a
natural candidate for the instrumental distribution in rejection
sampling; see also the last paragraph of Section~\ref{sec:conc}.

\section{A Two-Stage Modification and Comparative Study}\label{sec:2stage}

Babb et~al. (\citeyear{Babb98}) used EWOC to design a Phase~I trial to determine the
MTD, with $p=1/3$, of the antimetabolite 5-fluorouracil (5-FU) for the
treatment of solid tumors in the colon, when taken in conjunction with
fixed levels of the agents leucovorin (20 mg$/$m$^2$) and topotecan (0.5
mg$/$m$^2$). In this setting, a toxicity is considered a grade~4
hematologic or grade~3 or 4 nonhematologic toxicity within~2 weeks. As
mentioned above, EWOC involves specifying pre-trial a set
%
\begin{equation}\label{eq:levels}
x_{\min}=\lambda_1<\lambda_2<\cdots\le x_{\max}
\end{equation}
of possible dose values believed to contain the MTD, where $x_{\min}$
is taken as the starting value. Based on preliminary studies of 5-FU
given in conjunction with topotecan, a dose of $x_{\min}=140$ mg$/$m$^2$
of 5-FU was believed to be safe when given with~0.5 mg$/$m$^2$ of
topotecan. Also, a previous trial concluded that the MTD of 5-FU was
425 mg$/$m$^2$ when administered without topotecan, so $x_{\max}$ was
taken to be~425 mg$/$m$^2$ since 5-FU has been observed to be more toxic
when given with topotecan than alone. The two-parameter logistic
model~(\ref{1}) was chosen based on previous experience with the
agents, and uniform prior distributions over $[x_{\min}, x_{\max}]$
and\break $[0,0.2]$ were chosen for the MTD and the probability $F_\theta
(x_{\min})$, respectively. A~feasibility bound of $\omega=0.25$ was
chosen for EWOC and $p=1/3$. We compare EWOC and other previous designs
with the rollout (abbreviated by ROLL) of EWOC and the Hybrid~1 design
described in Section~\ref{sec:sim} in this setting with $n=24$.

To give a feel for the computational time required for the EWOC, ROLL
and Hybrid~1 designs, on a desktop personal computer with a 2.66 GHz
processor, the simulation of a single $n=24$ run of the ROLL design
with the EWOC base design took 49 minutes, whereas the Hybrid~1 design
took 0.4 seconds and EWOC took 0.12~seconds. The Hybrid~1 design is
computationally much simpler than ROLL since it does not perform
rollouts of a base design, but rather calculates its dose via (\ref
{eq:hyb_des}), where $m_k$ is the EWOC dose and $\ell_k$ is the
sequential $c$-optimal learning design. So although the interpolation
function (\ref{eq:epsk}) is derived from data gathered by ROLL during
its rollouts as described in \ref{sec:hyb_des}, the Hybrid~1 design
has computational time on the order of EWOC and the learning
design~$\ell_k$, even though the computational time required for ROLL
is large.

\begin{table*}
\caption{Risk, bias and RMSE of the final MTD estimate,
DLT rate and MTD overdose rate (OD), with SEs in parentheses, of
various designs}\label{table:2stage}
\begin{tabular*}{\tablewidth}{@{\extracolsep{4in minus 4in}}lccccc@{}}
\hline
\multicolumn{1}{@{}l}{\textbf{Design}}&\multicolumn{1}{c}{\textbf{Risk}}&\multicolumn{1}{c}{\textbf{Bias}}
&\multicolumn{1}{c}{\textbf{RMSE}}&\multicolumn{1}{c}{\textbf{DLT}}&\multicolumn{1}{c}{\textbf{OD}}\\
\hline
ROLL& 0.81 (0.01)& $-$0.069 (0.002) & 0.126 (0.022) &27.68\% (1.70\%) &
29.17\% (1.87\%) \\
\\
Hybrid 1&0.92 (0.03)&$-$0.075 (0.003)&0.128 (0.028)&24.68\% (0.86\%)&23.48\% (0.68\%)\\
\\
EWOC& 1.13 (0.01)& $-$0.076 (0.003) & 0.138 (0.024) &26.17\% (0.98\%) &
19.69\% (0.89\%) \\
\\
CRM& 1.65 (0.01)& \phantom{$-$}0.037 (0.003) & 0.118 (0.021)&36.37\% (1.10\%) &
62.69\% (1.10\%) \\
\\
$c$-opt& 1.71 (0.01)& \phantom{$-$}0.060 (0.003) & 0.126 (0.022) &23.44\% (0.95\%)
&12.42\% (0.74\%) \\
\\
$D$-opt& 1.96 (0.02)& $-$0.084 (0.006) & 0.143 (0.023) &13.55\% (0.31\%)
& \phantom{0}3.78\% (0.17\%) \\
\\
Wu& 1.77 (0.04)& \phantom{$-$}0.038 (0.009) & 0.122 (0.045)&23.40\% (0.54\%) &
40.25\% (0.77\%) \\
\\
SA& 1.52 (0.02)& \phantom{$-$}0.063 (0.003) & 0.131 (0.022)&22.39\% (0.93\%) &
35.56\% (0.40\%) \\
\\
$3+3_{10}$& 1.87 (0.01)& \phantom{$-$}0.060 (0.003) & 0.138 (0.024)&17.06\% (0.84\%)
& \phantom{0}0.85\% (0.21\%) \\
\\
$3+3_{20}$& 2.19 (0.02)& \phantom{$-$}0.070 (0.002) & 0.161 (0.025)&14.11\% (0.81\%)
& \phantom{0}0.75\% (0.29\%) \\
\hline
\end{tabular*}
\end{table*}

\subsection{A Comparative Study}\label{sec:comp}

Table \ref{table:2stage} first lists Bayesian designs, followed by
non-Bayesian designs that include Wu's~(\citeyear{Wu85}) design,
stochastic approximation (Lai and Robbins, \citeyear{Lai79}) and two \mbox{3-plus-3} dose
escalation designs. The first \mbox{3-plus-3}, denoted by $3+3_{10}$, uses 10
uniformly-spaced dose levels in $[x_{\min},x_{\max}]=[140,425]$. The
second uses 20~uniformly-spaced dose levels and is denoted by
$3+3_{20}$. Besides EWOC and its rollout ROLL, the Bayesian designs
include CRM, the constrained $D$-optimal design (abbreviated by
$D$-opt) of Haines et~al. (\citeyear{Haines03}) with constraint $\varepsilon=0.05$ and the
unconstrained sequential Bayesian $c$-optimal design (abbreviated by
$c$-opt) with $c$ being the vector $(0,1)^{\mathrm{T}}$. The prior density is
assumed to be uniform:
\begin{eqnarray}\label{eq:unifprior}
\pi(\rho,\eta)&=&[q(x_{\max}-x_{\min})]^{-1}\nonumber
\\[-8pt]\\[-8pt]
&&{}\cdot1\{(\rho,\eta
)\in[0,q]\times[x_{\min},x_{\max}]\}\nonumber
\end{eqnarray}
with $q=0.2$, where $1(A)$ denotes the indicator of a set~$A$. The
values of $(\rho,\eta)$ were generated from the prior distribution
(\ref{eq:unifprior}).

The performance of these designs is first evaluated in terms of the
global risk~(\ref{8}), in which we use the squared error~$g(\widehat
{\eta},\eta) =(\widehat{\eta}-\eta)^2$ for the MTD
estimate~$\widehat{\eta}=\widehat{\eta}(x_i,y_i,\ldots,x_n,y_n)$.
We then evaluate
performance exclusively in terms of the bias and root mean squared
error~(RMSE) of $\widehat{\eta}$ without taking into consideration the
risk to current patients, noting that the $c$- and $D$-optimal designs
focus on errors of post-trial parameter estimates. Finally, since
safety of the patients in the trial is the primary concern of
traditional 3-plus-3 designs, performance is also evaluated in terms of
the DLT rate and the probability of overdose (i.e., dose level
exceeding the MTD). Each result in Table \ref{table:2stage} is based on
2000 simulations.

\begin{figure*}[t]
\vspace*{4pt}
\includegraphics{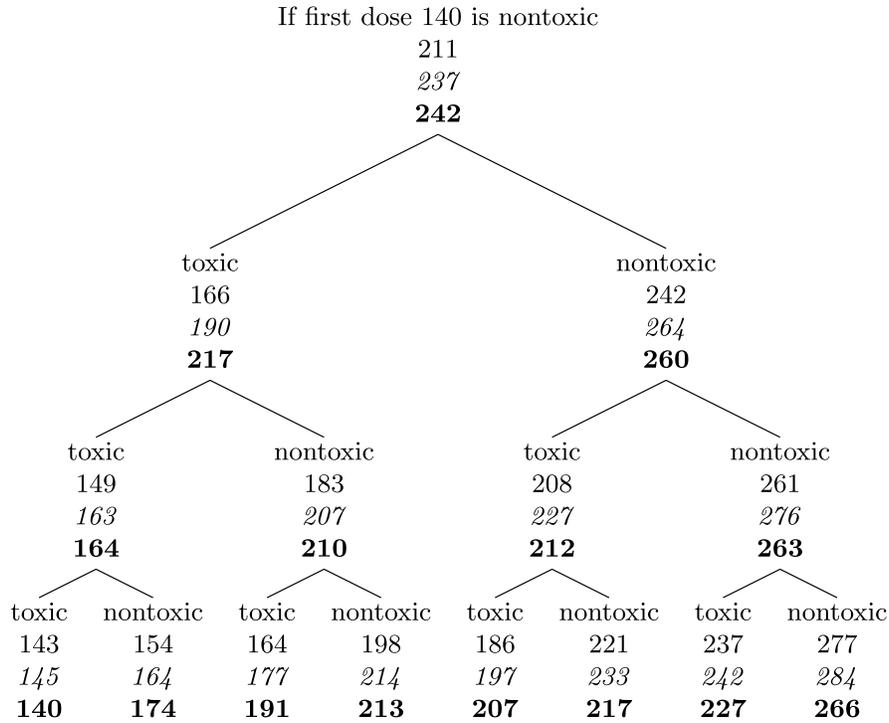}

\caption{The first five dose levels given by EWOC, ROLL (in italics)
and Hybrid~1 (in bold) in the 5-FU trial.}
\label{fig:tree}
\end{figure*}

The results in Table \ref{table:2stage} show that the effects of considering
the ``future'' patients is large, with ROLL and Hybrid 1 substantially
reducing the global risk from the myopic designs: in the case of ROLL,
about 30\% from EWOC, 35\% from CRM, and more from the 3-plus-3, $c$-
and $D$-opt, SA and Wu designs. Although ROLL has somewhat smaller
global risk than Hybrid~1, it is computationally much more expensive,
as noted above. The results for the 3-plus-3 designs show that they are
highly sensitive to the choice of $\lambda_1,\lambda_2,\ldots$ in
(\ref{eq:levels}). The $3+3_{10}$ design, using 10 uniformly-spaced
levels in $[x_{\min},x_{\max}]$, performs perhaps surprisingly well,
as it even has smaller risk than $D$-opt, which suffers from
substantial under-dosing due to its overdose constraint of $\varepsilon=0.05$.
This seems to be because the number~10 of dose levels was a fortuitous
choice given the parameter values and sample size of this study,
allowing the $3+3_{10}$ design to escalate to near the MTD in most
cases. However, there is often little information about the appropriate
number of doses and scale before a\break Phase~I cancer trial begins, and
when dose levels are chosen over a less fortuitous range or on too fine
a scale, as with the $3+3_{20}$ design, the majority of doses can end
up being administered at levels far below the therapeutic range near
the MTD. We emphasize that all of these designs are being evaluated
using the EWOC loss function in (\ref{7}) which, in particular for
CRM, differs from its associated loss function; using the CRM loss
function in (\ref{7}) results in CRM having smaller global risk than
EWOC, but again ROLL (with CRM as its base design) yields smaller
global risk than both and the relationship between the other designs
remains roughly unchanged.

In terms of MTD estimation accuracy, CRM and Wu have the smallest RMSE,
closely followed by $c$-opt and ROLL; CRM and Wu also have the smallest
absolute bias. It is interesting to note that the designs which
explicitly account for the asymmetric underdose/overdose relationship,
that is, ROLL, EWOC and $D$-opt (through its overdose
constraint~$\varepsilon
$), are negatively biased, while the others all have positive bias.

In terms of safety, $D$-opt and the $3+3$ designs have the smallest DLT
and OD rates, but in view of their large risk values, this safety comes
at the cost of low doses that are nontherapeutic. CRM has high
estimation accuracy and moderate risk, but also the largest DLT and OD
rates because of its symmetric loss function. The remaining designs,
ROLL, EWOC, Wu and SA, all have comparable DLT and OD rates, but their
risk values suggest that the magnitude of the overdoses in Wu and SA
are larger than EWOC, which, in turn, has larger overdoses than ROLL.

Of particular concern in phase~I trials is coherence of the design
(Cheung, \citeyear{Cheung05}), that is, whether the next patient will be
given a
higher dose if the current patient experiences a toxicity, and a lower
dose if the current patient does not. While a theoretical investigation
of the coherence of the ROLL and Hybrid designs is beyond our scope
here, as an illustrative example Figure~\ref{fig:tree} lists the first five
doses given by EWOC, ROLL and Hybrid~1 in the 5-FU trial setting,
assuming a nontoxic response to the first dose of $x_{\min}=140$. Note
that coherence is exhibited by all three designs in this example.

\begin{table*} 
\caption{Risk, bias and RMSE of the final MTD estimate,
DLT rate and MTD overdose rate (OD), with SEs in parentheses, of
various designs with the MTD fixed at the lower 15th percentile of the
misspecified prior}\label{table:misspec}
\begin{tabular*}{\tablewidth}{@{\extracolsep{4in minus 4in}}lcccccc@{}}
\hline
\textbf{Design}&&\textbf{Risk}&\textbf{Bias}&\textbf{RMSE}&\textbf{DLT}&\textbf{OD}\\
\hline
ROLL&(a)&1.64 (0.02) &$-$0.031 (0.003) & 0.142 (0.025) & 30.32\% (1.03\%)
& 39.38\% (1.09\%) \\
&(b)&1.39 (0.02) &$-$0.025 (0.003) & 0.145 (0.026) & 27.41\% (1.00\%) &
33.39\% (1.05\%) \\
\\
Hybrid~1&(a)&1.82 (0.05)&$-$0.032 (0.002)&0.151 (0.027)&36.90\% (1.52\%)&42.31\% (1.56\%)\\
&(b)&1.69 (0.04)&$-$0.027 (0.003)&0.131 (0.036)&35.70\% (1.51\%)&41.11\%
(1.56\%)\\
\\
EWOC&& 2.29 (0.02) &$-$0.034 (0.003) & 0.155 (0.028) & 35.33\% (1.07\%) &
45.98\% (1.11\%) \\
\\
CRM&&
3.83 (0.02) &\phantom{$-$}0.037 (0.004) & 0.179 (0.032) & 44.18\% (1.11\%) & 65.12\%
(1.07\%) \\
\hline
\end{tabular*}
\end{table*}

\subsection{A Two-Stage Design} \label{sec:mod3+3}

When one may have concerns about the validity of the Bayesian
parametric model in this model-based approach, one can readily
incorporate the hybrid designs as the second stage of a two-stage
design. The first stage of such escalates the doses cautiously by using
a modified 3-plus-3 design. For the batches of 3 in the 3-plus-3
design, we propose to combine the nonparametric step-up/down approach
with a parametric model-based dose determining scheme, thereby checking
the parametric model to be used for model-based escalation in the
second stage. This modification of the traditional 3-plus-3 design uses
a specified set of dose levels~(\ref{eq:levels}). Set $d_1=\lambda
_1=x_{\min}$. In the $k$th group of 3~patients, 2 patients are treated
at the same dose $d_k=\lambda_j$ and 1 patient at the EWOC dose $m_k$,
computed given the doses and responses of the previous $3(k-1)$
patients. If no DLT occurs in the group of 3 patients, $d_{k+1}$ is
increased to $\lambda_{j+1}$. If 1 DLT occurs, $d_{k+1}$ stays the
same at $d_k=\lambda_j$. Otherwise, 2 or 3~DLTS have occurred, so the
trial is stopped if $d_k=x_{\min}$, and \mbox{otherwise} continues with
$d_{k+1}$ lowered to $\lambda_{j-1}$. (Alternatively, it may be
desired to stop when 3 toxicities occur, regardless of what $d_k$ was.)
The EWOC dose $m_{k+1}$ is updated when the process is repeated with
the next group of 3 patients. This process repeats until a certain
fraction of the total number $n$ of patients has been treated, provided
the trial has not been stopped at the first stage due to excess
toxicities. We have found from our simulation studies that switch-over
points around $n/3$ or $n/4$ seem to strike a balance between enough
time for conservative dose escalation and model checking during the
first stage, while leaving enough time for efficient dose escalation in
the second stage.

The benefit of a first stage of conservative dose escalation occurs
when, unlike in Table \ref{table:2stage}, the prior distribution of the MTD
is misspecified. For example, if the true MTD falls in the left tail of
the prior distribution of $\eta$, then the prior information about the
MTD is biased upward, which can cause overdoses. In this situation,
including an initial stage of modified dose escalation, like the
modified 3-plus-3 scheme, provides additional safety by refining the
prior to be more accurate when it begins to be used in the second
stage. Focusing on the CRM, EWOC, ROLL and Hybrid~1 designs, Table \ref
{table:misspec} contains the results of a simulation study that
considers a situation such as this, where the true MTD is the lower
15th percentile of the MTD's nominal uniform prior distribution on
$[x_{\min},x_{\max}]$. That is, the data are generated with $\eta$
fixed at the 15th percentile of $[x_{\min},x_{\max}]$ and $\rho$
uniformly distributed over $[0,q]$, with $q=0.2$ as in Table \ref
{table:2stage}. The nominal prior for $(\rho,\eta)$ used by the
Bayesian procedures in Table \ref{table:misspec} is (\ref{eq:unifprior}),
the same as in Table \ref{table:2stage}, as are the values of the other
parameters. To see the effects of the first stage of more conservative
dose escalation, the operating characteristics of ROLL are recomputed
using a first stage of length $n/4=6$; the dose levels~(\ref
{eq:levels}) used by the modified 3-plus-3 design are 10
uniformly-spaced levels in $[x_{\min},x_{\max}]=[140,425]$. Adding
this first stage to ROLL or Hybrid~1 substantially reduces the risk,
DLT and overdose rates, as shown in Table \ref{table:misspec}, in which (a)
refers to the case of $n=24$ dose levels without the modified 3-plus-3
first stage, and (b) refers to the two-stage design using a first stage
of length $n/4=6$ consisting of the modified 3-plus-3 design.

\section{Conclusion}\label{sec:conc}

Despite their shortcomings and the development of alternative Bayesian
approaches since 1990, conventional dose-escalation designs are still
widely used in Phase~I cancer trials because of the ethical issue of
safe treatment of patients currently in the trial. However, a Phase~I
design also has the goal of determining the MTD for a future Phase~II
cancer trial, and needs an informative experimental design to meet this
goal. Von~Hoff and Turner (\citeyear{VonHoff91}) have documented that the overall response
rates in Phase~I trials are low and that substantial numbers of
patients are treated at doses that are retrospectively found to be
nontherapeutic. Eisenhauer et~al. (\citeyear{Eisenhauer00}), page~685, have pointed out that
``with a plethora of molecularly defined antitumor targets and an
increasingly clear description of tumor biology, there are now more
antitumor candidate therapies requiring Phase~I study than ever,'' and
that ``unless more efficient approaches are undertaken, Phase~I trials
may be a rate-limiting step in the process of evaluation of novel
anticancer agents.'' The hybrid designs in the previous section were
motivated by developing one such ``more efficient'' approach.

Hybrid designs with simple interpolation functions, refined through
iterated rollouts and regression, can be implemented by using simple
look-up tables for the parameters in (\ref{eq:f1}), and thus can be
relatively simple to use for clinicians. Given computer packages to
compute the standard myopic and learning designs, practitioners can use
a look-up table for the values $\varepsilon_k$ in (\ref{14}) as a function
of the relative posterior standard deviation $\nu_{k-1}/\nu_0$. For
given values of the prior parameters $x_{\min}, x_{\max}, p, q$ and
$\omega$, a computer package can generate this look-up table, which
can be used at every stage of the trial. We are in the process of
developing open source software for this purpose.

Tighiouart, Rogatko and Babb (\citeyear{Tighiouart05}) have shown how
Markov chain Monte Carlo (MCMC) can be used to compute the posterior
distribution of $(\rho,\eta)$ when the prior distribution is
supported on $[0,q]\times[x_{\min},\infty)$, extending the model
considered above\break where the support of $\eta$ is bounded above by
$x_{\max}$. They note that priors in this class with a negative
correlation structure between $\rho$ and $\eta$ result in an EWOC
design with comparable accuracy for estimating the MTD but lower DLT
and OD rates, relative to its performance for priors supported on
$[0,q]\times[x_{\min},x_{\max}]$. As noted in Section~\ref{sec:2stage}, a
two-stage design can easily address the higher DLT and OD rates caused
by misspecifications of such priors. On the other hand, even without a
cautious first stage, the above and other generalizations of the prior
of $(\rho,\eta)$ can be seamlessly incorporated into our hybrid
design. In fact, the model $M_4$ of Tighiouart et~al. (\citeyear{Tighiouart05}), which has
been shown to perform well in their simulation studies, has a
left-truncated, hierarchical normal prior distribution on $\eta$, so
the rejection sampling approach in the last paragraph of Section \ref
{sec:sim} can be applied here by using, say, the exponential
distribution as the instrumental distribution, since its tails are
upper bounds of those of the normal distribution. We can therefore
still use the Monte Carlo approach laid out at the end of Section \ref
{sec:sim}.

\section*{Acknowledgments}

Bartroff's work was supported by NSF Grant DMS-0907241
 and Lai's work was supported by NSF Grant DMS-0805879.

\end{document}